\documentclass{svproc}

\usepackage[version=3]{mhchem}
\usepackage[printonlyused, nolist]{acronym}
\usepackage{braket}
\usepackage{graphicx}
\usepackage{dcolumn}
\usepackage{bm}
\usepackage{subcaption}
\usepackage{placeins}
\usepackage{url}
\usepackage{amsmath,amssymb}
\usepackage{braket}

\begin{acronym}
    \acro{asf}[ASF]{Active Space Finder}
    \acro{ch2}[CH\textsubscript{2}]{methylene}
    \acro{dev2-svp}[dev2-SVP]{Development 2\textsuperscript{nd} generation - Split Valence Polarization}
    \acro{hea}[HEA]{Hardware-Efficient Ansatz}
    \acro{hf}[HF]{Hartree-Fock}
    \acro{h2}[H\textsubscript{2}]{hydrogen molecule}
    \acro{h4}[H\textsubscript{4}]{hydrogen chain}
    \acro{fci}[FCI]{full Configuration Interaction}
    \acro{hea}[HEA]{Hardware-Efficient Ansatz}
    \acro{hva}[HVA]{Hamiltonian Variational Ansatz}
    \acro{lih}[LiH]{lithium hydride}
    \acro{nisq}[NISQ]{Noisy Intermediate-Scale Quantum}
    \acro{pyscf}[PySCF]{Python-based Simulations of Chemistry Framework}
    \acro{qaoa}[QAOA]{Quantum Approximate Optimization Algorithm}
    \acro{qpe}[QPE]{Quantum Phase Estimation}
    \acro{scf}[SCF]{self-consistent field}
    \acro{sto3g}[STO-3G]{Slater-type orbitals with 3 Gaussian functions}
    \acro{tvha}[tVHA]{truncated Variational Hamiltonian Ansatz}
    \acro{ucc}[UCC]{Unitary Coupled Cluster}
    \acro{uccsd}[UCCSD]{Unitary Coupled Cluster with Single and Double excitations}
    \acro{vha}[VHA]{Variational Hamiltonian Ansatz}
    \acro{gd}[GD]{Gradient Descent}
    \acro{vqa}[VQA]{Variational Quantum Algorithm}
    \acro{vqe}[VQE]{Variational Quantum Eigensolver}
    \acro{nelder}[NM]{Nelder-Mead Algorithm}
    \acro{bfgs}[BFGS]{Broyden-Fletcher-Goldfarb-Shanno Algorithm}
    \acro{cmaes}[CMAES]{Covariance Matrix Adaptation Evolution Strategy}
    \acro{cobyla}[COBYLA]{Constrained Optimization By Linear Approximations}
    \acro{pso}[PSO]{Particle Swarm Optimization}
    \acro{slsqp}[SLSQP]{Sequential Least Squares Programming}
    \acro{spsa}[SPSA]{Simultaneous Perturbation Stochastic Approximation}
    \acro{ilshade}[iL-SHADE]{Improved Success-History Based Parameter Adaptation for Differential Evolution with Linear Population Size Reduction}
    \acro{sacauchy}[SA Cauchy]{Simulated Annealing Cauchy}
    \acro{HS}[HS]{Harmony Search}
    \acro{SOS}[SOS]{Symbiotic Organisms Search}
    \acro{iSOMA}[iSOMA]{improved SOMA}
    \acro{bp}[BP]{Barren Plateau}
\end{acronym}
\begin{document}
\mainmatter

\title{Reliable Optimization Under Noise in Quantum Variational Algorithms}

\titlerunning{Reliable Optimization Under Noise in VQAs}

\author{Vojt\v{e}ch Nov\'{a}k\inst{1,2} \and
Silvie Ill\'{e}sov\'{a}\inst{6} \and
Tom\'{a}\v{s} Bezd\v{e}k\inst{3} \and
Ivan Zelinka\inst{2,4} \and
Martin Beseda\inst{5}
}

\authorrunning{V. Nov\'{a}k et al.}

\tocauthor{Vojt\v{e}ch Nov\'{a}k, Silvie Ill\'{e}sov\'{a}, Tom\'{a}\v{s} Bezd\v{e}k, Ivan Zelinka, Martin Beseda}

\institute{
Department of Computer Science, Faculty of Electrical Engineering and Computer Science,\\
VSB - Technical University of Ostrava, Ostrava, Czech Republic\\
\email{vojtech.novak.st1@vsb.cz}
\and
IT4Innovations National Supercomputing Center,\\
VSB - Technical University of Ostrava, 708 00 Ostrava, Czech Republic
\and
Department of Mathematics, TUM School of Computation, Information and Technology, Technical University of Munich, Garching bei München, Germany
\and
Department of Informatics and Statistics, Marine Research Institute,\\
Klaipeda University, Lithuania
\and
Dipartimento di Ingegneria e Scienze dell’Informazione e Matematica,\\
Università dell’Aquila, Via Vetoio, I-67010 Coppito, L’Aquila, Italy
\and
Gran Sasso Science Institute, L’Aquila, Italy 
}

\maketitle

\begin{abstract}
The optimization of Variational Quantum Eigensolver is severely challenged by finite-shot sampling noise, which distorts the cost landscape, creates false variational minima, and induces statistical bias called winner's curse. We investigate this phenomenon by benchmarking eight classical optimizers spanning gradient-based, gradient-free, and metaheuristic methods on quantum chemistry Hamiltonians \ce{H2}, \ce{H4} chain, LiH (in both full and active spaces) using the truncated Variational Hamiltonian Ansatz. We analyze difficulties of gradient-based methods (e.g., SLSQP, BFGS) in noisy regimes, where they diverge or stagnate. We show that the bias of estimator can be corrected by tracking the \textit{population mean}, rather than the biased best individual when using population based optimizer. Our findings, which are shown to generalize to hardware-efficient circuits and condensed matter models, identify adaptive metaheuristics (specifically CMA-ES and iL-SHADE) as the most effective and resilient strategies. We conclude by presenting a set of practical guidelines for reliable VQE optimization under noise, centering on the co-design of physically motivated ansatz and the use of adaptive optimizers.
\end{abstract}

\section{Introduction}

Optimization under stochastic noise presents a universal challenge in numerical and physical sciences. In black-box optimization, where the objective function is accessed only through noisy evaluations, random fluctuations can distort the apparent topology of the loss landscape. Additive Gaussian noise leads to a statistical phenomenon known as the \emph{winner’s curse}, where the lowest observed energy or cost is biased downward relative to the true expectation value due to random fluctuations \cite{Andrews2019Inference,Rakshit2017NoisyEASurvey}. This bias results in the optimizer prematurely accepting a spurious minimum as the global optimum, preventing the discovery of better, higher accuracy solutions. When applied to quantum variational algorithms, this stochastic effect acquires deeper physical implications. In the context of the \ac{vqe} \cite{cerezo2021variational}, the cost function is the variational expectation value
\begin{equation}
C(\boldsymbol{\theta}) = \langle \psi(\boldsymbol{\theta}) | \hat{H} | \psi(\boldsymbol{\theta}) \rangle ,
\label{cost}
\end{equation}
where \(\ket{\psi(\boldsymbol{\theta})} = U(\boldsymbol{\theta}) \ket{0}\) is the variational ansatz prepared by the parameterized unitary \(U(\boldsymbol{\theta})\) acting on an easy to prepare reference state \(\ket{0}\). The cost function should be bounded by the variational principle from below by the true energy of the ground state, \(C(\boldsymbol{\theta}) \geq E_0\). However, in practice, we can only estimate the expectation value of the cost function up to a finite precision determined by the number of measurement shots, \(N_{\mathrm{shots}}\). The corresponding estimator reads
\begin{equation}
\bar{C}(\boldsymbol{\theta}) = C(\boldsymbol{\theta}) + \epsilon_{\text{sampling}},
\label{cost_noisy}
\end{equation}
where \(\epsilon_{\text{sampling}}\) is a zero-mean random variable, typically modeled as Gaussian noise, \(\epsilon_{\text{sampling}} \sim \mathcal{N}(0, \sigma^2 / N_{\mathrm{shots}})\). This sampling noise can lead to apparent violations of the variational bound, yielding \(\bar{C}(\boldsymbol{\theta}) < E_0\), a phenomenon known as \emph{stochastic variational bound violation}. 

This appears as false variational minima, illusory states that seem better than the true ground state, but arise solely from statistical fluctuations \cite{Wierichs2021NoiseMinima,McClean2019VQATheory}. These minima can mislead the optimizer, inducing premature convergence or divergence depending on the search strategy. This effect is further compounded by the presence of a \emph{noise floor}, a finite lower limit in achievable precision defined by the sampling variance of the observable \cite{McClean2019VQATheory}.

Another fundamental trainability limitation of VQE arises from the \emph{\ac{bp}} phenomenon \cite{Larocca2025BPReview}. In the presence of BPs, gradients of the loss function or equivalently expectation differences concentrate exponentially around their mean with increasing the number of qubits, rendering the landscape effectively flat and featureless. The loss function can be expressed in operator form as
\begin{equation}
\ell_{\boldsymbol{\theta}}(\rho, O) = \mathrm{Tr}[\rho(\boldsymbol{\theta}) O],
\end{equation}
where \(\rho(\boldsymbol{\theta}) = U_{\boldsymbol{\theta}}(\rho)\) and $\rho = \ket{\psi}\bra{\psi}$ is the density matrix. Equation~(3) can be interpreted using the Hilbert-Schmidt inner product 
\(\langle A, B\rangle = \mathrm{Tr}[A^\dagger B]\), so that 
\(\ell_{\boldsymbol{\theta}}(\rho, O) = \langle \rho(\boldsymbol{\theta}), O\rangle\) 
represents the overlap between two operators viewed as vectors in the operator space \(B(\mathcal{H}_0)\) of dimension \(4^n\). Optimization can thus be viewed as the anti-alignment of these two exponentially large vectors, which far exceeds the Hilbert space dimension \(2^n\). As discussed in the \ac{bp} review \cite{Larocca2025BPReview}, this exponential operator space structure fundamentally limits trainability, producing exponentially vanishing gradients and effectively concealing the direction of improvement under any finite sampling precision.

Recent work has broadened the VQE framework to include advanced variants such as the State-Averaged Orbital-Optimized VQE (SAOOVQE), allowing ground and excited-state descriptions under realistic quantum noise conditions \cite{beseda2024state,illesova2025transformation,saoovqebenchmarking,rajamani2025equiensembledescriptionsystematicallyoutperforms}. The reliability of variational optimization under noise is likewise relevant beyond quantum chemistry, influencing quantum machine learning \cite{illesova2025qmetric,illesova2025importance,gupta2022quantum}, condensed-matter modeling \cite{vorwerk2022quantum,PhysRevA.111.022437}, and emerging practical uses such as medical diagnostics \cite{novak2025predicting,novakpropensity} and software testing \cite{trovato2025preliminary,evolvingcircuits}.

In this work, we extend our previous investigation of optimization strategies for the \ac{tvha}~\cite{Illesova2025VHA} by broadening the scope to multiple ansatz types and physical models. Our aim is to provide the reader with a clearer intuition for how finite-shot sampling noise reshapes variational energy landscapes across both problem-inspired (\ac{tvha}) and hardware-efficient (TwoLocal) circuits. Using molecular (\ce{H2}, \ce{H4}, LiH with full and active space), Ising, and Fermi–Hubbard systems, we visualize how smooth convex basins deform into rugged multimodal surfaces as noise increases. These visualizations, together with systematic benchmarks, expose how different optimizers respond to stochastic distortions in the loss landscape.

We benchmarked gradient-based, gradient-free, and evolutionary optimizers under identical noise conditions. Gradient-based methods (gradient descent, \ac{slsqp}~\cite{boggs1995sequential,kraft1988software}, \ac{bfgs}~\cite{liu1989}) rapidly lose stability once the cost curvature approaches the sampling noise level, producing unreliable gradient and Hessian estimates. Population-based methods (\ac{cmaes}, PSO~\cite{eberhart1995particle,marini2015particle}) remain functional through implicit averaging but suffer from the \emph{winner’s curse}, showing false improvements unless high-shot reevaluation is used~\cite{Andrews2019Inference,Rakshit2017NoisyEASurvey}. This motivates statistically grounded acceptance rules and adaptive resampling in noisy variational optimization.

Importantly, these effects are not specific to the \ac{tvha} or to quantum chemistry. Our complementary study on the 1D Ising and Fermi–Hubbard models using a hardware-efficient TwoLocal ansatz~\cite{Novak2025VQAmeta} reveals the same phenomena: gradient-based optimizers have difficulties converging, and finite-shot noise converts smooth basins into rugged, multimodal landscapes. Observing identical behaviors across two distinct ansatz classes (physically motivated \ac{tvha} versus generic TwoLocal) and across domains (molecular chemistry versus condensed matter) suggests that these stochastic distortions and optimizer failures are \emph{generic features} of noisy VQE, situating our results within a broader framework of noisy hybrid optimization~\cite{McClean2019VQATheory,Wierichs2021NoiseMinima}.

The \ac{tvha} ansatz itself mitigates barren plateaus through compact, physically motivated structure and initialization near the Hartree--Fock state, ensuring polynomially decaying gradients~\cite{Illesova2025VHA,park2024hva}. This distinguishes it from hardware-efficient circuits and makes ansatz design a concrete \emph{noise-mitigation strategy}.

In summary, Section~1 introduces the motivation and theoretical background, emphasizing stochastic variational bound violations and barren plateaus in VQE. Section~2 outlines the analytical and numerical methods, including stochastic bias modeling, benchmark Hamiltonians, and the construction of the \ac{tvha} and TwoLocal circuits, we also relate optimizer step sizes to the Hessian spectrum, showing that when steps exceed local curvature scales, noise dominates updates. Section~3 presents results that compare classical optimizers with varying noise levels, while Section~4 summarizes results from all systems studied, the role of structured ansatz design, and the advantages of adaptive evolutionary methods. Appendix~A lists the implementation details and reproducibility settings.

\section{Methods}

This section outlines the analytical and numerical techniques used to characterize the impact of finite sampling on variational quantum optimization. We first quantify how measurement noise introduces \emph{stochastic selection bias} and generates false minima, establishing a statistical framework for evaluating bias magnitude and noise floors. We then analyze how this stochasticity reshapes the underlying VQE optimization landscape, examining its interplay with circuit expressivity, ansatz structure, and optimizer behavior under realistic shot noise.

\subsection{Stochastic selection bias and false minima under finite sampling}

Variational quantum algorithms (VQAs) rely on parameterized quantum circuits (PQCs), also called ansatz, to generate states of the form
\begin{equation}
\ket{\psi(\boldsymbol{\theta})} = U(\boldsymbol{\theta}) \ket{0},
\end{equation}
where the unitary transformation can be expressed as a sequence of parameterized layers
\begin{equation}
U(\boldsymbol{\theta}) = \prod_{i=1}^{L} U_i(\theta_i),
\end{equation}
with \(\boldsymbol{\theta} = (\theta_1, \dots, \theta_L)\) representing trainable parameters. These parameters can be interpreted as effective evolution times under the corresponding generators, for example \(U(\theta) = e^{-i\theta H}\).

The objective of the algorithm is to minimize the cost function defined in Eq.~\eqref{cost}. However, as shown in Eq.~\eqref{cost_noisy}, in practice the expectation value can only be estimated up to finite precision determined by the number of measurement shots, \(N_{\text{shots}}\). For a molecular Hamiltonian decomposed into Pauli strings $\hat{H} = \sum_k c_k P_k$, the intrinsic variance follows
\begin{equation}
\mathrm{Var}[C(\boldsymbol{\theta})] = \langle \hat{H}^2 \rangle - \langle \hat{H} \rangle^2 ,
\end{equation}
with covariance contributions arising within commuting groups used to reduce the measurement count.

Each function evaluation (FE) in the optimization corresponds to one noisy estimate $\bar{C}(\boldsymbol{\theta}_i)$ computed from $N_{\text{shots}}$ samples. The optimization algorithm then selects the smallest of $K$ such draws:
\begin{equation}
\bar{C}_{\min} = \min_{i \in \{1,\dots,K\}} \bar{C}(\boldsymbol{\theta}_i).
\end{equation}
Even though $\mathbb{E}[\epsilon_{\text{sampling}}]=0$, the statistical minimum $\bar{C}{\min}$ is biased downward because the minimum of noisy random variables systematically overestimates improvement. Following extreme-value statistics \cite{Andrews2019Inference}, the expected bias magnitude scales as 
\begin{equation}
\mathbb{E}[\bar{C}{\min}] - C_{\min} \approx -\sigma_{\text{noise}}\sqrt{2\log K}
\end{equation}
where $K$ is the number of function evaluations. In practice, correlated measurements reduce the number of statistically independent samples, effectively lowering $K$ and hence the true bias \cite{Andrews2021WinnersCurse}. For large evaluation counts, one or more noisy estimates will almost certainly fall several standard deviations below the mean, creating false minima; improvements smaller than about four to five noise standard deviations should therefore be treated as statistically insignificant.

The noise floor for each configuration was computed as
\begin{equation}
\sigma_{\text{noise}} = \sqrt{\frac{\bar{\sigma}^2}{N_{\text{shots}}}},
\end{equation}
where $\bar{\sigma}^2$ denotes the mean variance of all Hamiltonian evaluations during optimization. For shot counts below a few thousand, the noise floor already exceeds the typical energy improvement step, implying that observed minima below the ground-state reference are spurious statistical outliers rather than physical convergence.

\subsection{Model Hamiltonians and Ansatzes}

We benchmarked our optimization strategies using two classes of Hamiltonians and corresponding variational ansatzes. The 1D Ising and 6-site Fermi–Hubbard models were simulated using the \textit{TwoLocal} ansatz, while molecular systems (\ce{H2}, \ce{H4} chain and \ce{LiH}) used tVHA. 

The 1D Ising Hamiltonian without external field is defined as
\begin{equation}
H = -\sum_{i=1}^{n-1} \sigma_z^{(i)} \sigma_z^{(i+1)},
\end{equation}
where $\sigma_z^{(i)}$ denotes the Pauli-$Z$ operator acting on qubit $i$, and $n$ is the total number of qubits representing the spin chain \cite{lieb1961two}. 

The 6-site Fermi–Hubbard model combines kinetic and on-site interaction terms,
\begin{equation}
H = -t \sum_{\langle i,j\rangle,s} (c_{i,s}^\dagger c_{j,s} + c_{j,s}^\dagger c_{i,s}) + U \sum_i n_{i,\uparrow} n_{i,\downarrow},
\end{equation}
where $c_{i,s}^\dagger$ ($c_{i,s}$) are fermionic creation (annihilation) operators for spin $s \in \{\uparrow,\downarrow\}$ at site $i$, $\langle i,j\rangle$ denotes nearest-neighbor pairs of lattice sites, $n_{i,s} = c_{i,s}^\dagger c_{i,s}$ is the number operator, and $t=U=1$ \cite{hubbard1963electron}. Both the Ising and Hubbard models were used to evaluate optimizer performance on structured yet nontrivial quantum landscapes.

For molecular systems, we used the \ac{tvha}, which mitigates the limitations of traditional quantum chemistry ansatzes such as UCC that often lead to deep, non-commuting circuits and exponential parameter scaling. The \ac{tvha} is systematically derived from the molecular Hamiltonian
\begin{align}
H &= \sum_{ij} h_{ij} a_i^\dagger a_j 
  + \frac{1}{2} \sum_{ijk\ell} g_{ijk\ell} a_i^\dagger a_j^\dagger a_k a_\ell,
\end{align}
where $a_i^\dagger$ ($a_i$) are fermionic creation (annihilation) operators acting on orbital $i$, $h_{ij}$ are one-electron integrals representing kinetic and nuclear-attraction contributions, and $g_{ijk\ell}$ are two-electron integrals describing electron–electron interactions. A truncation parameter $p$ limits weaker non-Coulomb terms by retaining only the dominant subset that preserves a chosen fraction of the total interaction strength. The \ac{vha} and its truncated variants have been extensively analyzed in the context of entanglement, optimization, and expressivity \cite{wiersema2020exploring,park2024hva,truncatedVHA2025,Illesova2025VHA}.

\subsection{VQE optimization landscape: geometry, noise, and step-size stability}

\begin{figure*}[htpb]
  \centering
  \setlength{\tabcolsep}{2pt}
  \renewcommand{\arraystretch}{0}
  \begin{tabular}{ccc}
    \includegraphics[width=.31\textwidth]{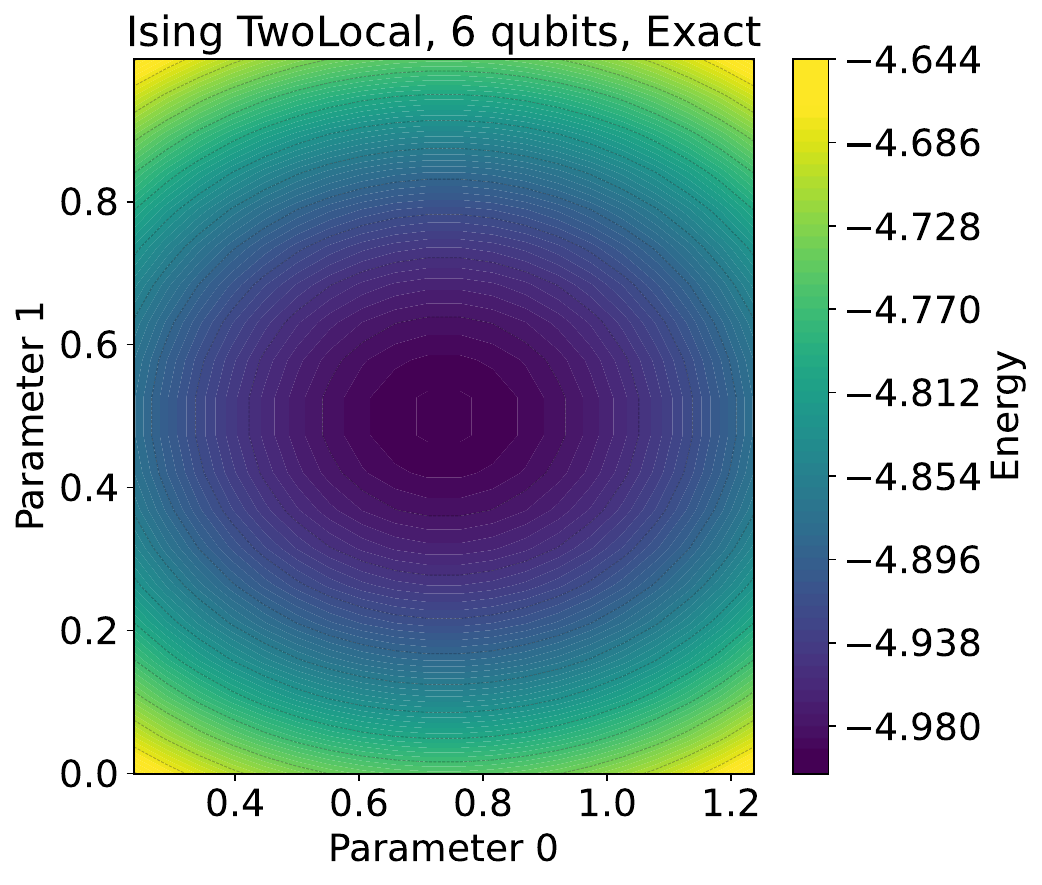} &
    \includegraphics[width=.31\textwidth]{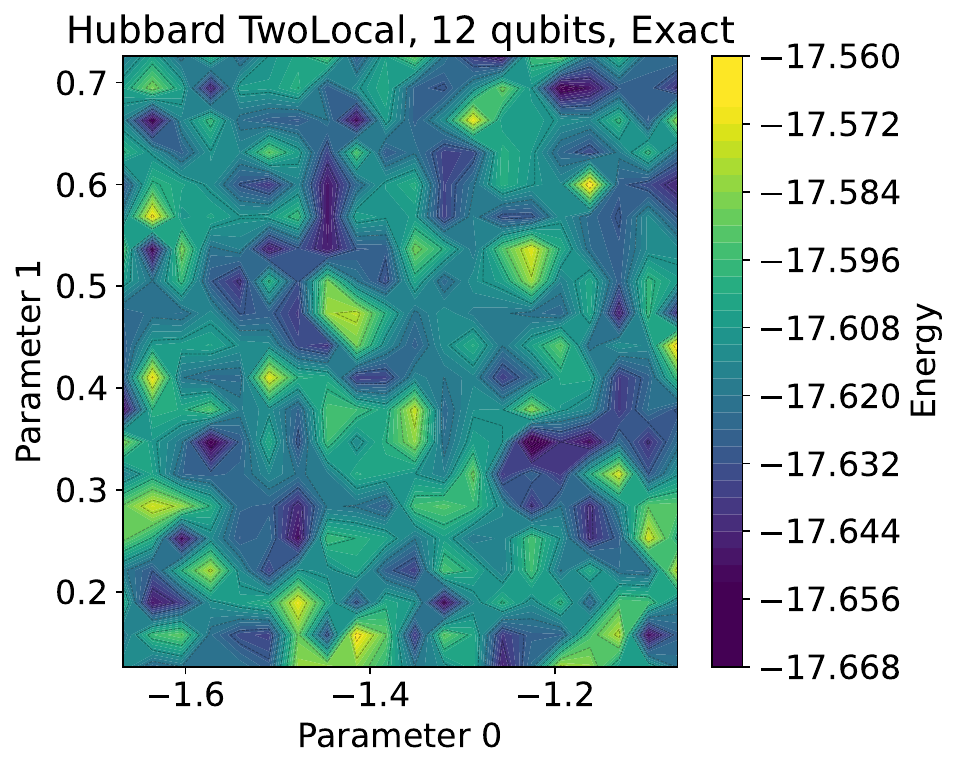} &
    \includegraphics[width=.31\textwidth]{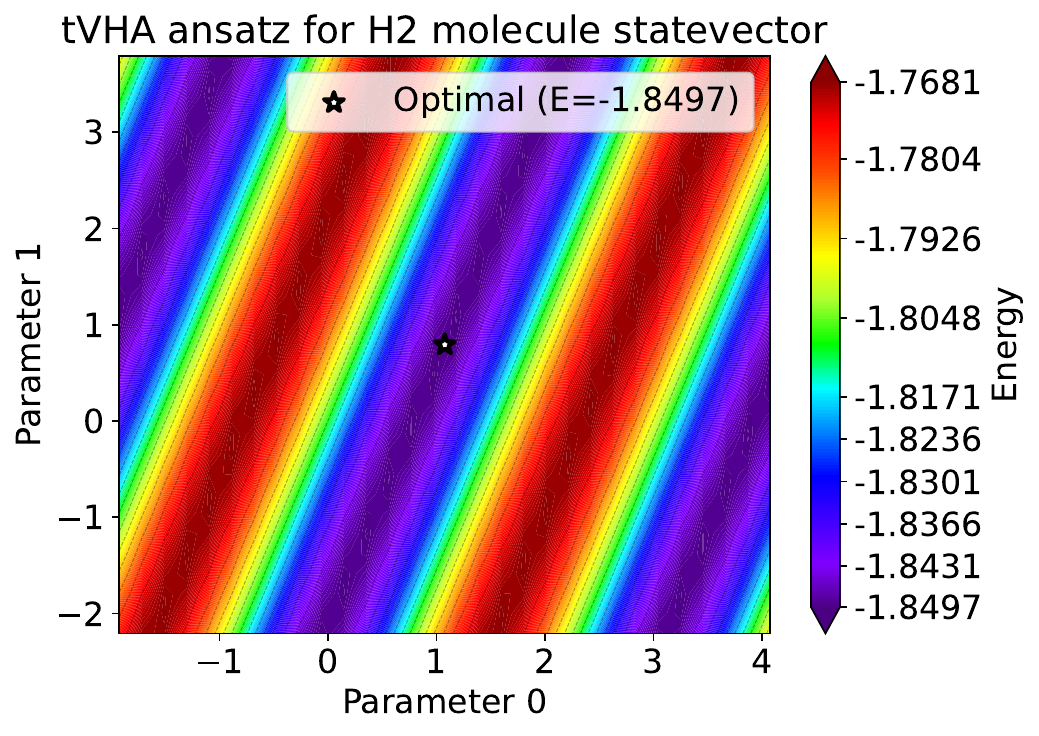} \\

    \includegraphics[width=.31\textwidth]{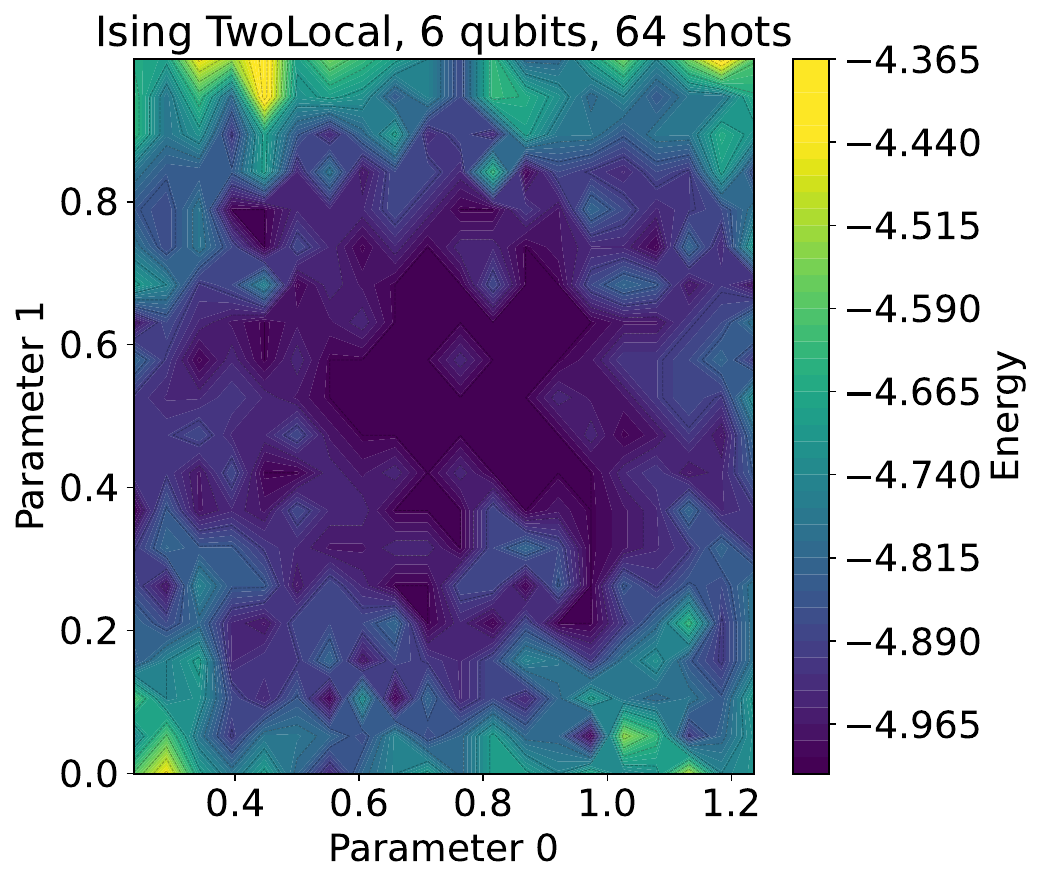} &
    \includegraphics[width=.31\textwidth]{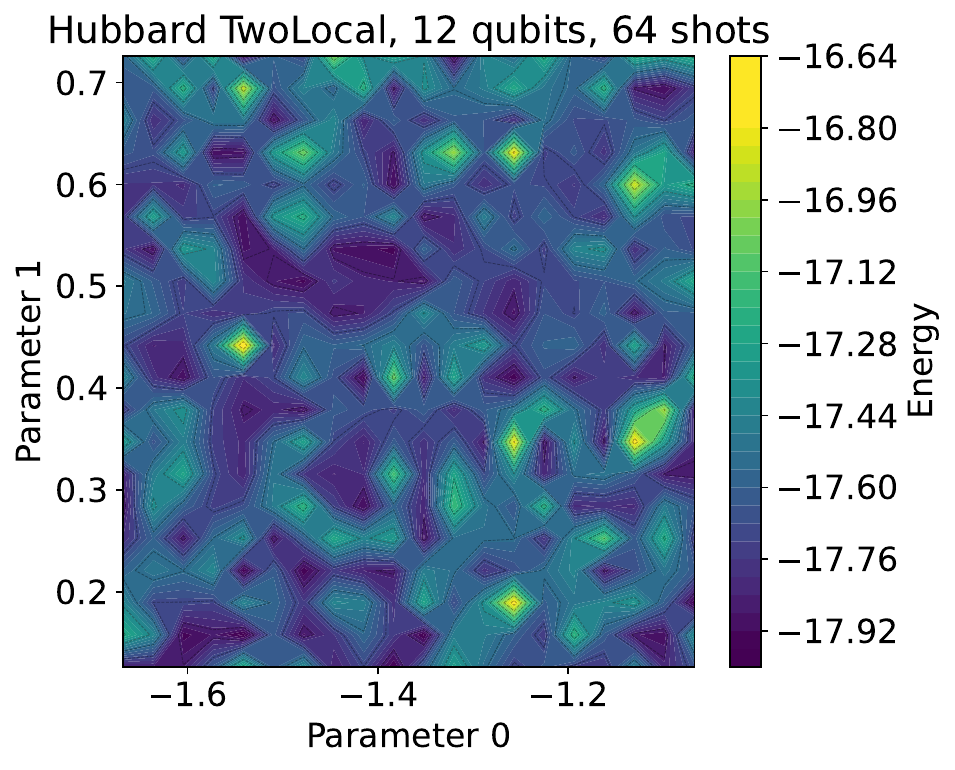} &
    \includegraphics[width=.31\textwidth]{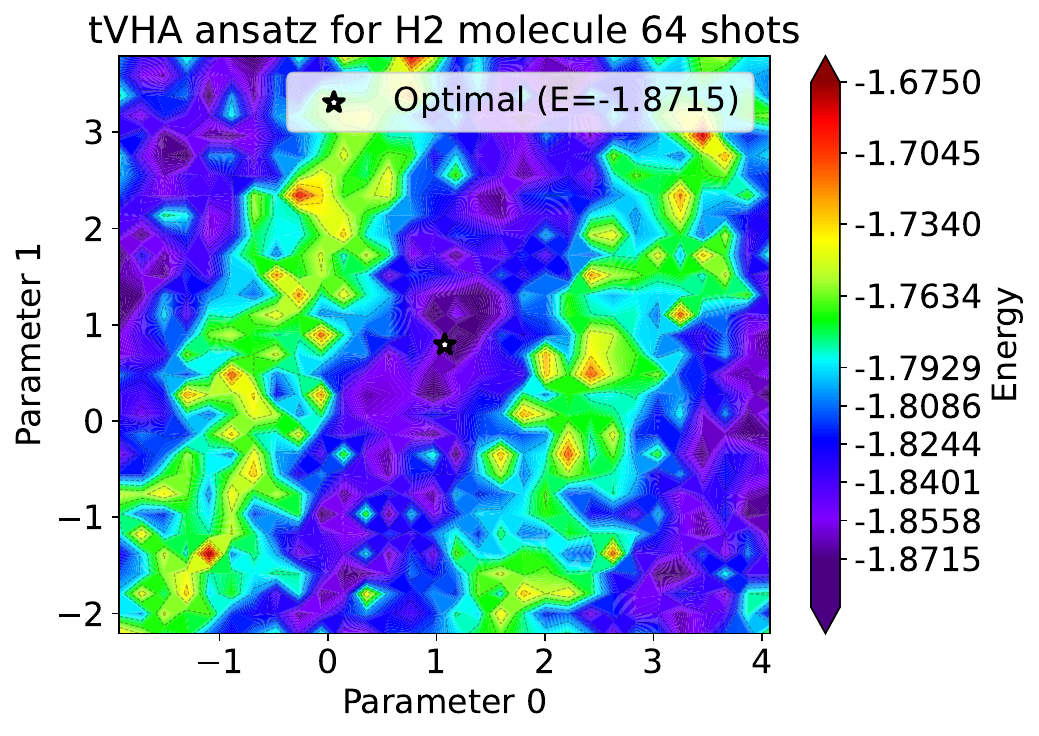} \\

    \includegraphics[width=.31\textwidth]{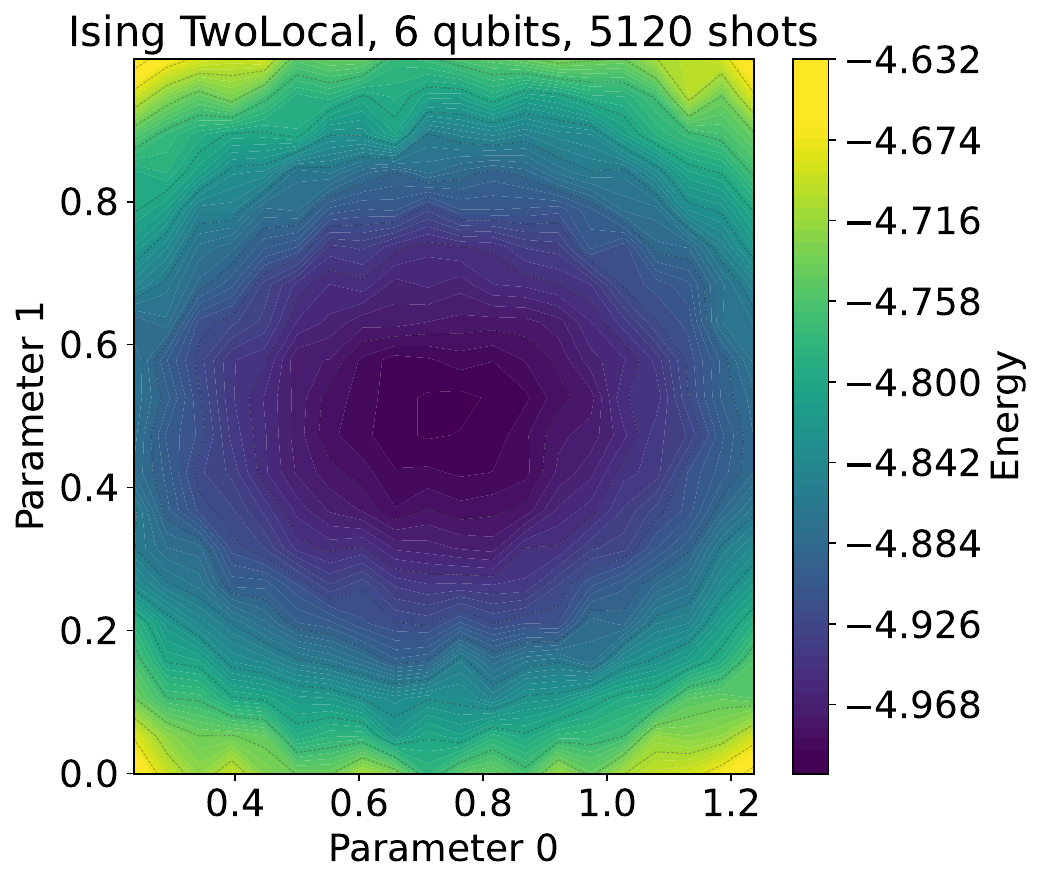} &
    \includegraphics[width=.31\textwidth]{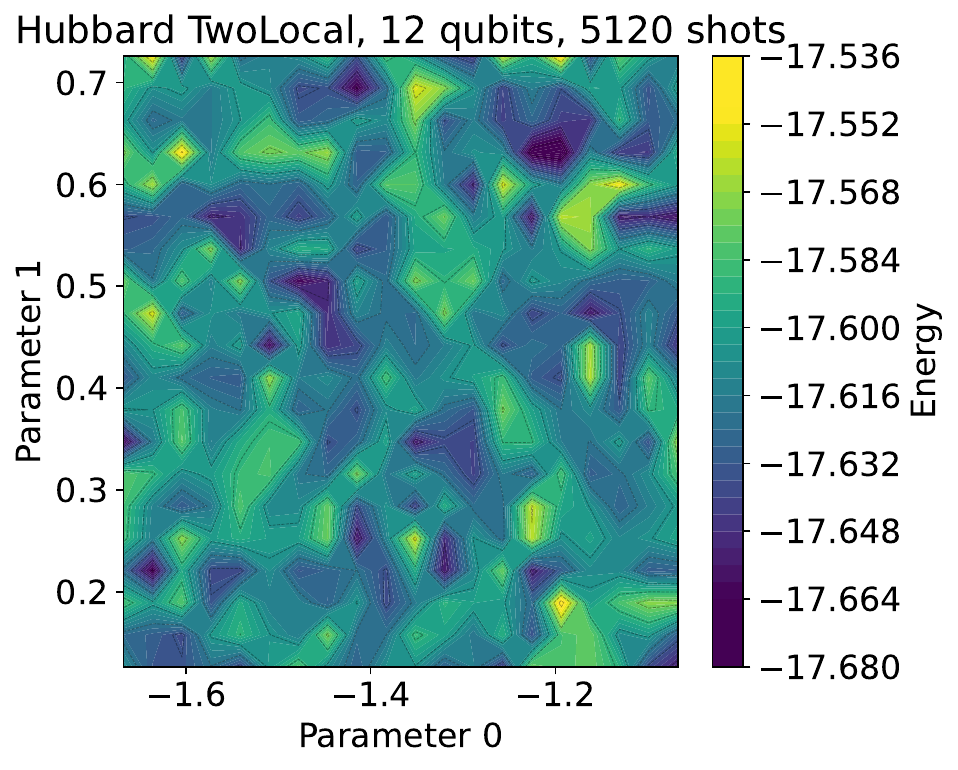} &
    \includegraphics[width=.31\textwidth]{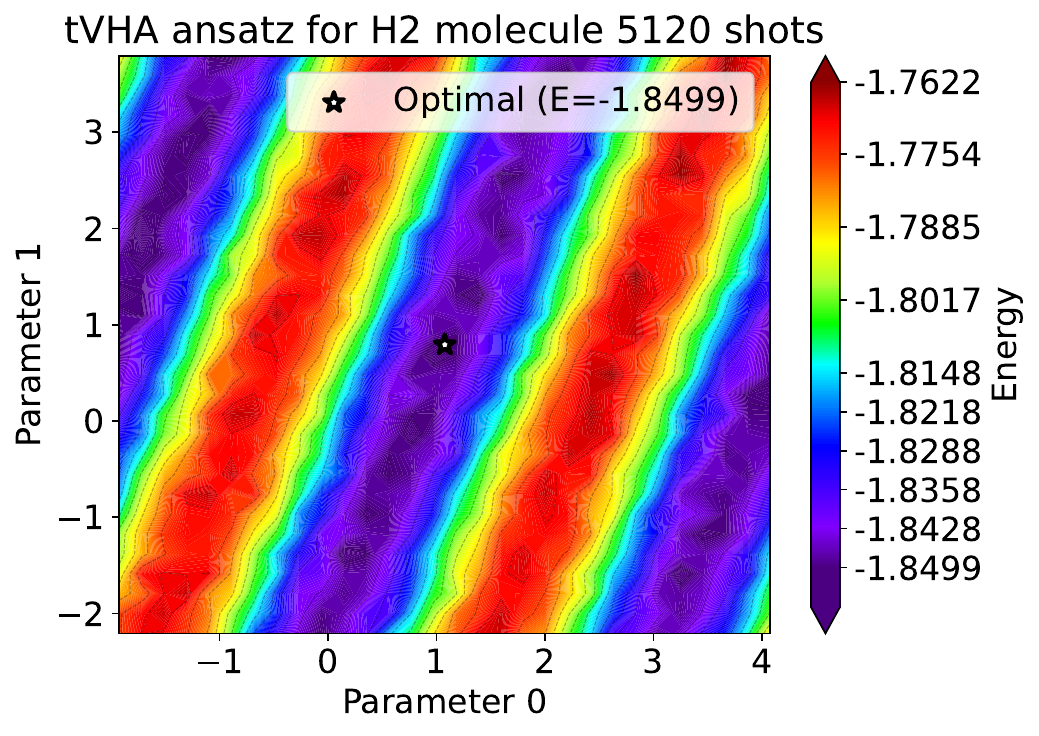} \\

    \multicolumn{3}{c}{
      \includegraphics[width=.31\textwidth]{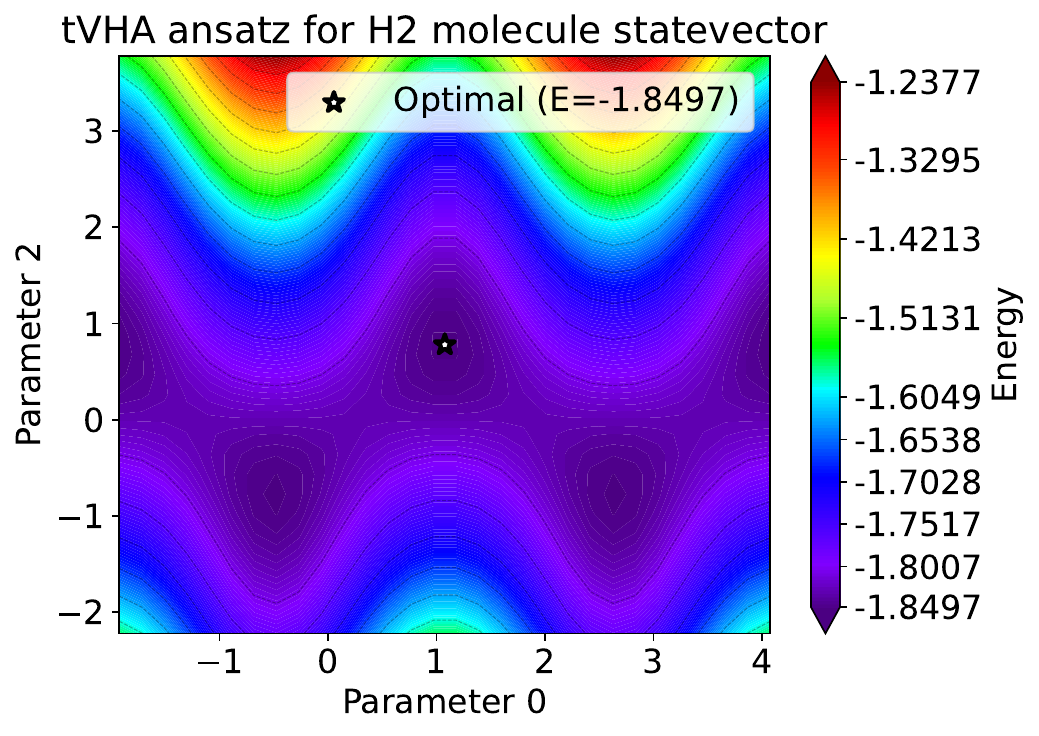} \hspace{2pt}
      \includegraphics[width=.31\textwidth]{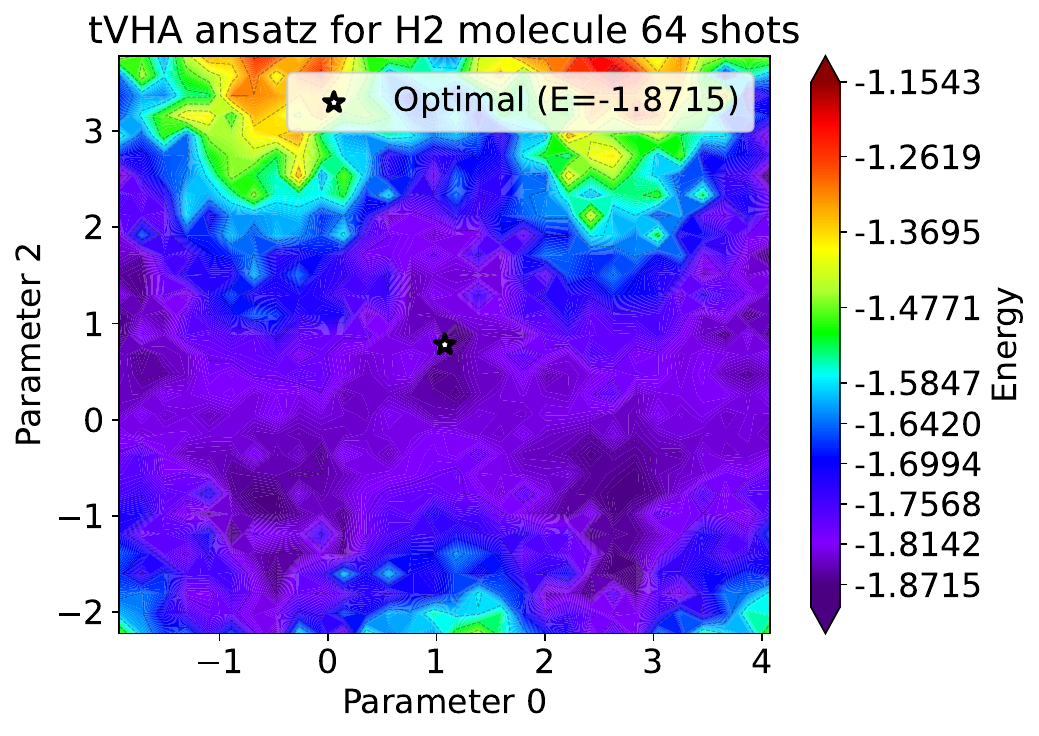} \hspace{2pt}
      \includegraphics[width=.31\textwidth]{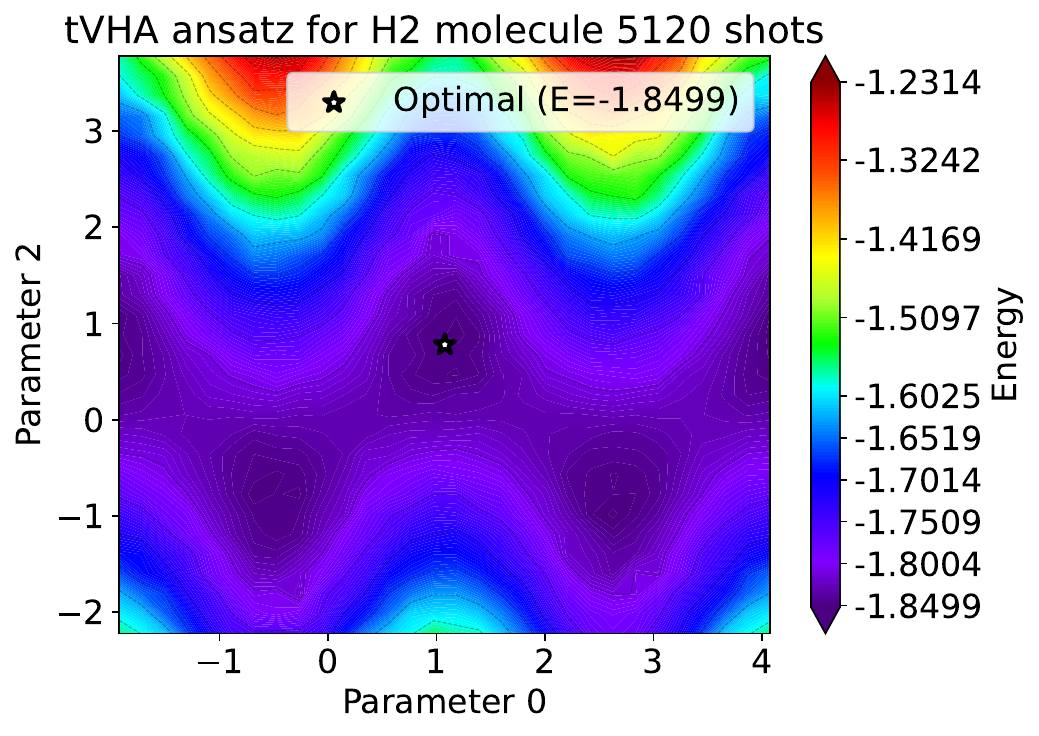}
    }
  \end{tabular}
  \caption{
  Combined comparison of energy landscapes across models and noise levels.
  Columns correspond to the Ising model (left), Hubbard model (center), and \ce{H2} system (right).
  Rows correspond to statevector/noiseless (top), 64–512-shots (second), 5120–$6{\times}1024$-shots (third), and alternative \ce{H2} parameter slices (bottom).
  }
  \label{fig:combined_landscapes_4x3}
\end{figure*}

The geometry of the variational landscape governs VQE trainability and optimizer robustness. In the noiseless regime, structured ansatzes such as the \ac{tvha} form smooth, locally convex basins around the Hartree–Fock initialization, while hardware-efficient circuits (e.g., TwoLocal) introduce mild curvature irregularities. Strongly correlated models like the six-site Hubbard system are inherently multimodal even under exact simulation \cite{BonetMonroig2021PerformanceComparison,Wang2020NoiseInduced}. Figure~\ref{fig:combined_landscapes_4x3} compares these topologies, showing two-dimensional parameter slices where only selected parameters vary and others are fixed at their optimized values.

The connection between local curvature and optimizer stability can be made explicit by analyzing the Hessian. 
Near an optimum \(\boldsymbol{\theta}_\star\), the cost function can be expanded as
\begin{equation}
C(\boldsymbol{\theta}) \approx C(\boldsymbol{\theta}_\star) 
+ \tfrac{1}{2} (\boldsymbol{\theta} - \boldsymbol{\theta}_\star)^{\top} 
\boldsymbol{H} (\boldsymbol{\theta} - \boldsymbol{\theta}_\star),
\end{equation}
where \(\boldsymbol{H} = \nabla^2 C(\boldsymbol{\theta}_\star)\) denotes the local Hessian.  
Unless stated otherwise, all vector norms and inner products are taken with respect to the standard Euclidean (\(\ell_2\)) geometry.

For gradient descent with step size \(\eta\), the linearized iteration
\begin{equation}
\boldsymbol{\theta}_{k+1} = \boldsymbol{\theta}_k - \eta \nabla C(\boldsymbol{\theta}_k)
\end{equation}
is locally stable when the spectral radius of \((\boldsymbol{I} - \eta \boldsymbol{H})\) is less than one. 
For a symmetric positive definite Hessian, this implies
\begin{equation}
0 < \eta < \frac{2}{L},
\end{equation}
where \(L = \lambda_{\max}(\boldsymbol{H})\) is the Lipschitz constant of the gradient.  
The asymptotic convergence rate depends on the condition number 
\(\kappa = L / \mu\), with \(\mu = \lambda_{\min}(\boldsymbol{H})\).  
A large \(\kappa\) indicates strong anisotropy of the curvature and results in slow convergence along shallow directions.

When finite-shot sampling noise is present, the measured cost \(\bar{C}(\boldsymbol{\theta})\) becomes stochastic, 
and the corresponding gradient estimator satisfies
\begin{equation}
\nabla \bar{C}(\boldsymbol{\theta}) 
= \nabla C(\boldsymbol{\theta}) + \boldsymbol{\xi}, 
\quad \mathbb{E}[\boldsymbol{\xi}] = 0, \quad 
\operatorname{Cov}[\boldsymbol{\xi}] = \Sigma,
\end{equation}
where \(\boldsymbol{\xi}\) represents the random gradient noise arising from finite-shot sampling, modeled as a zero-mean random vector with covariance matrix \(\Sigma\) that quantifies the sampling-induced uncertainty in each gradient component.
Once the root-mean-square gradient noise exceeds the local signal scale,
\begin{equation}
\sqrt{\mathbb{E}[\|\boldsymbol{\xi}\|_2^2]} 
\gtrsim \|\boldsymbol{H}(\boldsymbol{\theta} - \boldsymbol{\theta}_\star)\|_2,
\end{equation}
the optimizer can no longer reliably detect the descent direction.  
In this regime dominated by noise, the apparent curvature fluctuates between iterations, 
so that a fixed step size \(\eta \approx 2/L\) alternates between being too small and too large, 
leading to oscillatory or divergent behavior.  
Stable operation therefore requires both a well-conditioned Hessian 
and gradient estimates whose variance is small compared to the curvature scale.

\begin{figure}[htpb] \centering \includegraphics[width=0.5\linewidth]{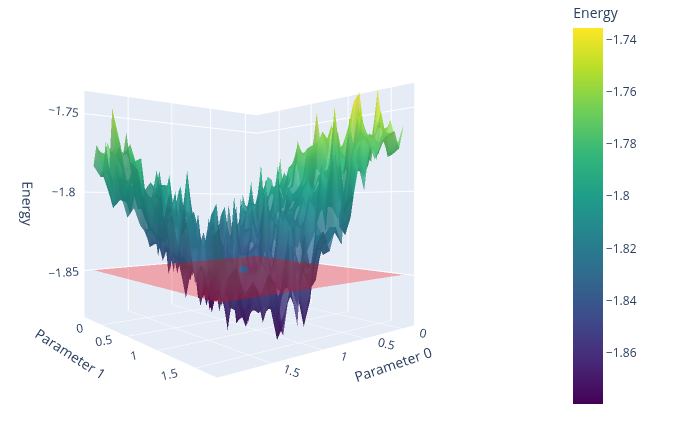} \caption{Energy landscape slice for noisy VQE estimation of \ce{H2} using 512 shots with \ac{tvha} ansatz. The red plane indicates the exact ground-state energy $E_0$; apparent dips below $E_0$ arise from sampling noise, violating the variational principle (false minima).} \label{fig:3d} \end{figure}

As illustrated in Figure \ref{fig:3d}, the energy landscape slice for the noisy VQE estimation of the H\(_2\) molecule, using 512 shots with the \ac{tvha} ansatz, highlights this issue. The red plane indicates the exact ground-state energy \(E_0\), but the apparent dips below \(E_0\) arise from the sampling noise, violating the variational principle. These false minima, which emerge due to stochastic fluctuations in the optimization process, emphasize the difficulty of avoiding variational violations in noisy quantum environments. The winner's curse bias exacerbates this issue, causing the optimizer to chase these fluctuations and engage in a random walk that reduces the effectiveness of the optimization process.

Deterministic curvature-based optimizers such as \ac{bfgs} and \ac{slsqp} rely on consistent curvature updates and monotonic objective decrease, so stochastic fluctuations often cause erratic or divergent steps \cite{Novak2025VQAmeta,BonetMonroig2021PerformanceComparison}. In contrast, stochastic and population-based strategies, such as SPSA \cite{Spall1998SPSA} and \ac{cmaes} \cite{Hansen2006CMAES} average over noisy evaluations, maintaining stability in rugged or noise-dominated landscapes, though they still benefit from periodic high-precision reevaluation of elite candidates.

\section{Results}

\begin{figure}[htpb]
    \centering
    \includegraphics[width=0.5\linewidth]{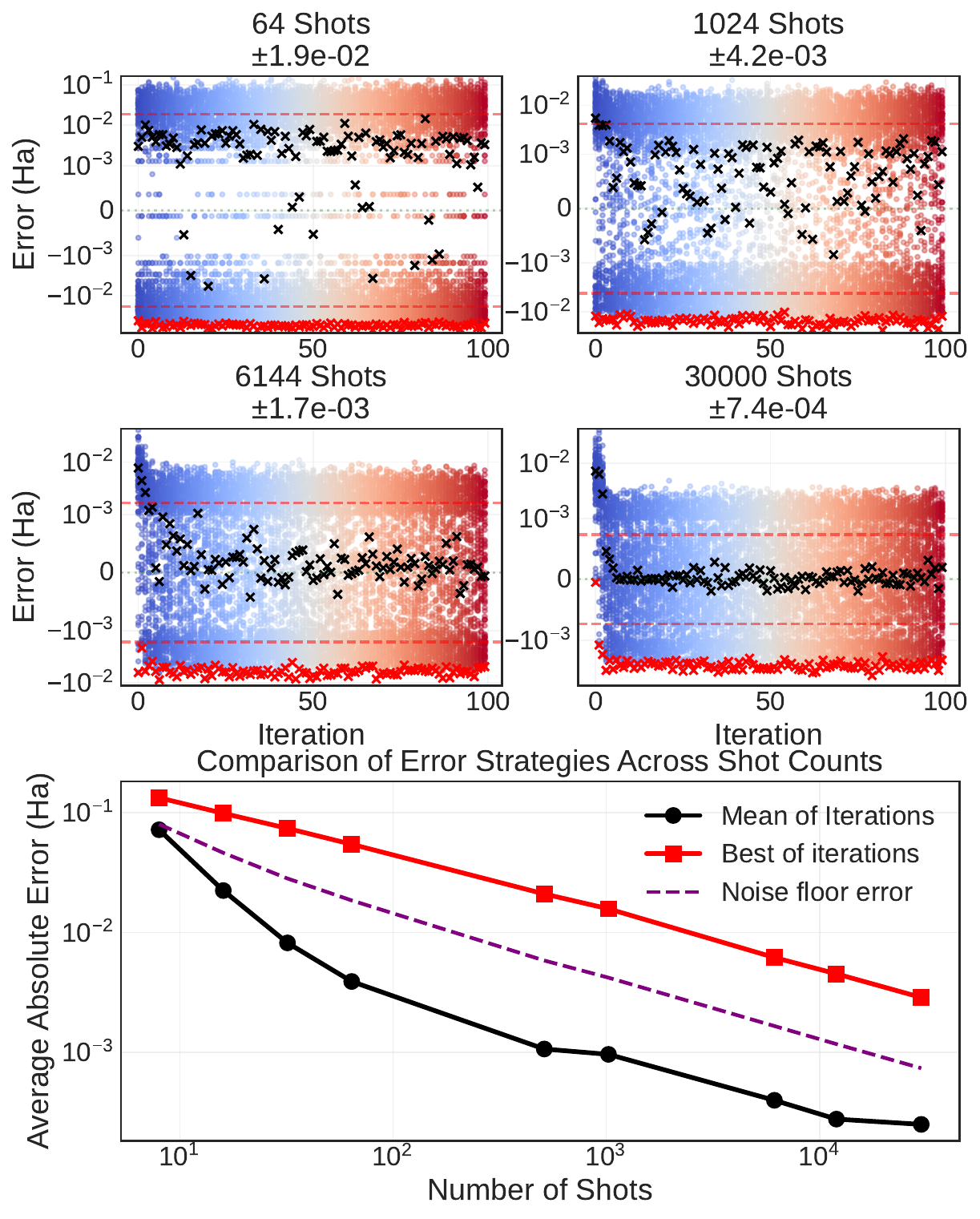}
    \caption{Energy error from all individuals for \ce{H2} using \ac{tvha}. Top: Optimization trajectories (colored points), iteration means (black crosses), best values (red crosses), and noise floors (red dashed lines). Bottom: Average absolute errors for mean-based (black), best-value (red), $\sigma_{\text{noise}}$ (purple).}
    \label{fig:floor}
\end{figure}

\begin{figure}[htpb]
    \centering
    \includegraphics[width=0.75\linewidth]{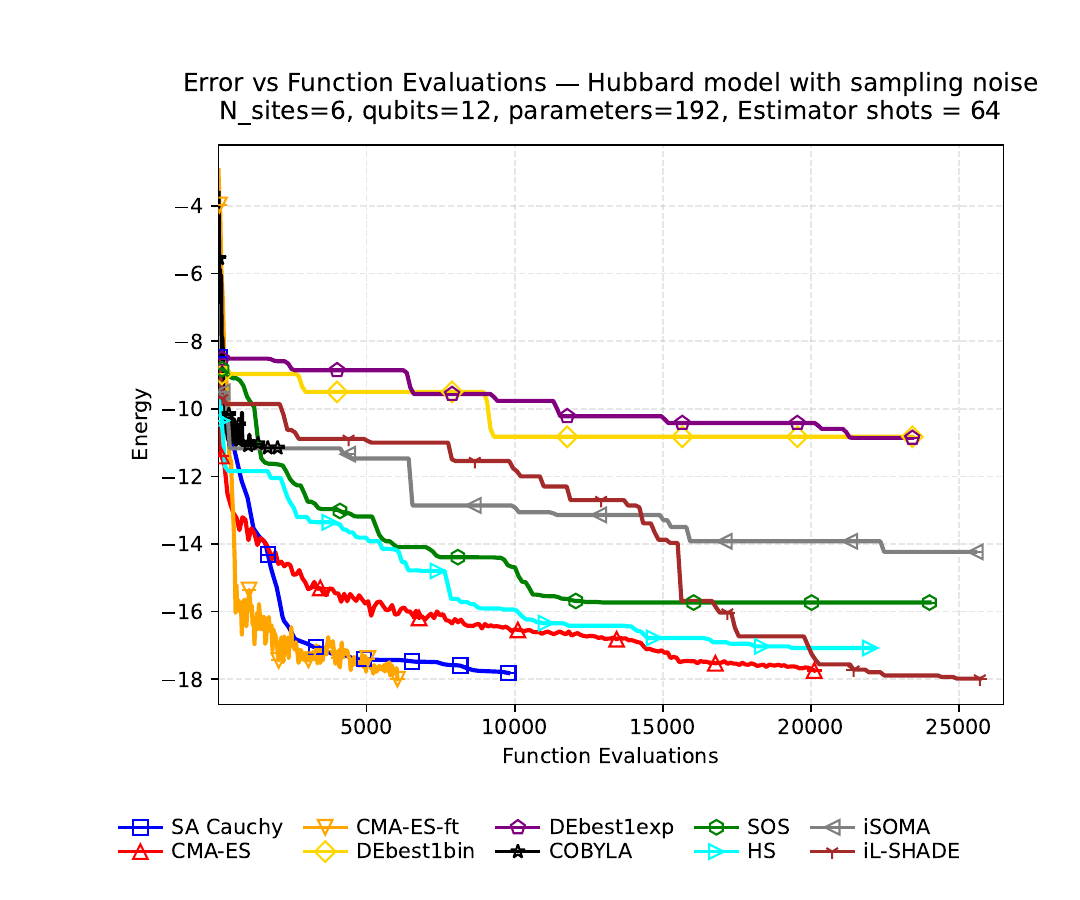}
    \caption{Convergence comparison for the six-site Hubbard model using 64-shot estimation. Adaptive metaheuristics (iL-SHADE and \ac{cmaes}) maintain stable progress where gradient-based and basic Differential Evolution variants stagnate.}
    \label{fig:hubbard}
\end{figure}

Here we show results for \ac{cmaes} optimizer in \ce{H2} system using \ac{tvha} under various noise settings in Fig. \ref{fig:floor}. In this part, we analyze the magnitudes of the noise floor $\sigma_{\text{noise}}$ and corresponding errors from tracking the lowest FE in iteration (Best of iterations) compared to averaging all FEs in the iteration (Mean of iterations). Next, we show results for the Ising model in Tab. \ref{tab:algorithm_performance_sorted}. This table shows how many FEs optimizers needed converge to optimum point in tolerance $1e-1$ (we show only successful optimizers). After that we present convergence curves of optimizers for the six-site Hubbard model in Fig. \ref{fig:hubbard}. Finally, we show convergence curves of optimizers using \ac{tvha} in \ce{H2}, \ce{H4} chain, \ce{LiH} full space and \ce{LiH} active space systems using random initializations in statevector (exact) and sampling noise simulations.

The results in Fig.~\ref{fig:floor} demonstrate that combining population-based optimization with averaging across individuals in each iteration significantly suppresses stochastic fluctuations in the energy estimates. The \emph{best of iterations} metric often falls spuriously below the true ground-state energy due to overfitting to random sampling noise, while the population-mean trajectory remains bounded by the empirical noise floor $\sigma_{\text{noise}}$ and follows the expected $1/\sqrt{N_{\text{shots}}}$ scaling. For the \ac{cmaes} optimizer, this effect is clearly observed across increasing shot numbers. At $N_{\text{shots}} = 64$, the mean absolute error is approximately $4 \times 10^{-3}$~Ha, compared to $5 \times 10^{-2}$~Ha for the best-sample metric and a corresponding noise floor of $1.9 \times 10^{-2}$~Ha. Increasing to $N_{\text{shots}} = 1024$ reduces these values to $1 \times 10^{-3}$~Ha (mean), $1.6 \times 10^{-2}$~Ha (best), and $4.2 \times 10^{-3}$~Ha ($\sigma_{\text{noise}}$). At $N_{\text{shots}} = 6144$, the mean and noise floor drop to $4 \times 10^{-4}$~Ha and $1.7 \times 10^{-3}$~Ha, respectively, and at $N_{\text{shots}} = 30000$ they converge further to $2.5 \times 10^{-4}$~Ha and $7.4 \times 10^{-4}$~Ha. This progression confirms that ensemble averaging yields nearly an order of magnitude smaller effective errors than tracking only the best individuals, turning sampling noise from a limiting factor into a statistically manageable uncertainty.

The results in Table~\ref{tab:algorithm_performance_sorted} demonstrate a clear scaling hierarchy in optimizer efficiency with increasing qubit number. 
\ac{cmaes} consistently achieves the fastest convergence across all system sizes, requiring fewer than $3.5\times10^3$ function evaluations even for nine qubits. 
The adaptive \ac{ilshade} follows closely, showing mild growth with system size and maintaining stable convergence within $6.6\times10^3$ evaluations. 
Both methods exhibit near linear scaling and robust performance, confirming their ability to adapt search distributions effectively in high-dimensional parameter spaces.

\begin{table}[htpb]
\centering
\caption{Algorithm performance comparison for Ising model without external magnetic field of five to nine qubits. Values represent the mean number of function evaluations (FEs) required for convergence across five independent runs.}
\small
\renewcommand{\arraystretch}{0.85}
\begin{tabular}{|l|r|r|r|r|r|}
\hline
Algorithm & 5Q & 6Q & 7Q & 8Q & 9Q \\
\hline
\ac{cmaes} & 1500 & 1800 & 2280 & 2700 & 3200 \\
SA Cauchy & 3500 & 4201 & 7000 & 8001 & 9451 \\
iL-SHADE & 3274 & 3333 & 4368 & 4374 & 6559 \\
HS & 1740 & 1726 & 5046 & 7066 & 10586 \\
DEbest1bin & 4460 & 8112 & 11676 & 14720 & 17136 \\
iSOMA & 5552 & 15263 & 22115 & 28144 & 33899 \\
SOS & 4880 & 13440 & 15320 & 32000 & 37843 \\
\hline
\end{tabular}
\label{tab:algorithm_performance_sorted}
\end{table}

In contrast, \ac{sacauchy} and \ac{HS} show moderate scalability, with function evaluations increasing rapidly beyond seven qubits. 
Their slower adaptation and limited population diversity lead to premature saturation as the landscape becomes more rugged. 
DEbest1bin, \ac{iSOMA}, and \ac{SOS} \cite{CHENG201498} perform significantly worse, exhibiting near exponential growth in required evaluations and frequent stagnation at higher dimensions. 
These results highlight the superior efficiency and scalability of adaptive metaheuristics, particularly \ac{cmaes} and iL-SHADE, for variational quantum optimization in expanding Hilbert spaces.

The convergence behavior for the six-site Fermi–Hubbard model, shown in Fig.~\ref{fig:hubbard}, reveals a pronounced separation between adaptive metaheuristics and conventional optimizers under 64-shot sampling noise. 
Among all tested methods, \ac{ilshade} and \ac{cmaes}-ft achieve energies closest to the exact ground state $E_0=-18$~Ha, reaching final values of $-17.984$ and $-17.998$~Ha, corresponding to absolute errors of $1.6\times10^{-2}$ and $2.7\times10^{-3}$~Ha, respectively. 
Both algorithms maintain smooth, monotonic convergence over the full trajectory, indicating stable adaptation of search distributions even in the presence of measurement noise.

The unmodified \ac{cmaes} and \ac{sacauchy} runs converge to slightly higher energies ($-17.745$ and $-17.818$~Ha), corresponding to residual errors of $2.6\times10^{-1}$ and $1.8\times10^{-1}$~Ha. 
Although these methods exhibit consistent early progress, their trajectories plateau prematurely as noise-induced fluctuations obscure curvature information. 
In contrast, \ac{HS} and \ac{SOS} display slower and less stable descent, saturating near $-17.08$ and $-15.73$~Ha (errors of $9.2\times10^{-1}$ and $2.27$~Ha), suggesting partial trapping in false minima amplified by sampling variance.

Local search and basic evolutionary schemes perform substantially worse. 
\ac{cobyla} terminates near $-11.16$~Ha, while both DEbest1bin and DEbest1exp stagnate around $-10.83$~Ha, corresponding to errors exceeding $7$~Ha. 
The \ac{iSOMA} variant, despite a larger evaluation budget, remains confined above $-14.23$~Ha, more than $3.7$~Ha$ $ above the ground state. 
Such outcomes reflect the inability of fixed-parameter heuristics to cope with the rugged, noise-deformed landscape characteristic of the Hubbard model.

Overall, Fig.~\ref{fig:hubbard} demonstrates that only adaptive metaheuristics achieve consistent and near-variational convergence in the noisy Hubbard regime, while deterministic and static search strategies terminate well above the true ground-state energy.

For \ce{H2}, the simplest system, all optimizers except gradient descent converge to near machine precision in noiseless simulations, with \ac{bfgs}, \ac{cmaes}, COBYLA, Nelder–Mead, and PSO reaching errors below $10^{-13}$–$10^{-14}$ Ha. Gradient descent and SPSA are slower, with final errors of $\sim 10^{-6}$ and $3\times 10^{-5}$ Ha, respectively. Under sampling noise, the spread across optimizers increases markedly. After high-shot reevaluation, gradient descent, SPSA, and \ac{cmaes} yield the lowest corrected mean errors, all near $4$–$7\times 10^{-4}$ Ha, while \ac{bfgs}, Nelder–Mead, and COBYLA degrade to the millihartree range ($2$–$4\times 10^{-3}$ Ha). PSO remains competitive ($\sim 6\times 10^{-4}$ Ha), showing resilience to noise despite its slower noiseless convergence. These results reveal that in \ce{H2}, all optimizers succeed in the noiseless limit, but under realistic sampling conditions, stochastic and population-based methods retain sub-millihartree accuracy, whereas deterministic gradient-based ones lose stability.

The increase in system complexity for \ce{H4} further separates optimizer performance. In statevector runs, \ac{bfgs}, COBYLA, and \ac{slsqp} achieve errors of $\sim 7\times 10^{-3}$ Ha, while \ac{cmaes} and Nelder–Mead converge more slowly, and gradient descent reaches $\sim 2.5\times 10^{-2}$ Ha. When sampling noise is introduced, the corrected high-shot results show that \ac{cmaes} gives the lowest mean error ($2.3\times 10^{-2}$ Ha), closely followed by PSO ($2.7\times 10^{-2}$ Ha) and gradient descent ($3.0\times 10^{-2}$ Ha). SPSA and COBYLA lie in the intermediate range ($3.5$–$3.8\times 10^{-2}$ Ha), while \ac{bfgs} and Nelder–Mead deteriorate strongly (above $4.5\times 10^{-2}$ Ha). Thus, for \ce{H4}, gradient-based methods lose their noiseless advantage, and only population-based and stochastic algorithms retain partial accuracy.

For the full \ce{LiH} molecule, the higher-dimensional ansatz amplifies these trends. In statevector simulations, \ac{bfgs}, \ac{cmaes}, COBYLA, Nelder–Mead, and SPSA reach errors near $4\times 10^{-3}$ Ha, while PSO and gradient descent are less precise ($9\times 10^{-3}$ and $2\times 10^{-2}$ Ha). Under noise, \ac{cmaes} achieves the best corrected mean error ($\sim 8\times 10^{-3}$ Ha), followed by PSO ($1.3\times 10^{-2}$ Ha) and SPSA ($1.6\times 10^{-2}$ Ha). All deterministic optimizers (\ac{bfgs}, COBYLA, Nelder–Mead) cluster near $2\times 10^{-2}$ Ha, consistent with reduced robustness to stochastic gradients. These results indicate that population-based and stochastic optimizers maintain sub-20 mHa accuracy even for larger systems, while deterministic gradient methods fail to scale effectively.

The \ce{LiH} active space demonstrates the benefit of orbital truncation, both in optimizer cost and noise resilience. In the noiseless setting, all algorithms achieve sub-millihartree accuracy, with \ac{cmaes}, PSO, Nelder–Mead, and SPSA converging below $2\times 10^{-4}$ Ha, and \ac{bfgs} and COBYLA reaching $\sim 2\times 10^{-4}$ Ha. Under sampling noise, high-shot reevaluation reveals \ac{cmaes} as the best performer ($2.6\times 10^{-4}$ Ha), followed by SPSA ($3.9\times 10^{-4}$ Ha), PSO ($4.2\times 10^{-4}$ Ha), and Nelder–Mead ($5.3\times 10^{-4}$ Ha). COBYLA and \ac{bfgs} yield slightly higher errors ($8\times 10^{-4}$–$1.1\times 10^{-3}$ Ha), while gradient descent deteriorates to several millihartree and \ac{slsqp} diverges beyond $3\times 10^{-1}$ Ha. Active space reduction thus improves stability by roughly an order of magnitude relative to the full system.

\begin{figure}[htpb]
    \centering
    \includegraphics[width=1\linewidth]{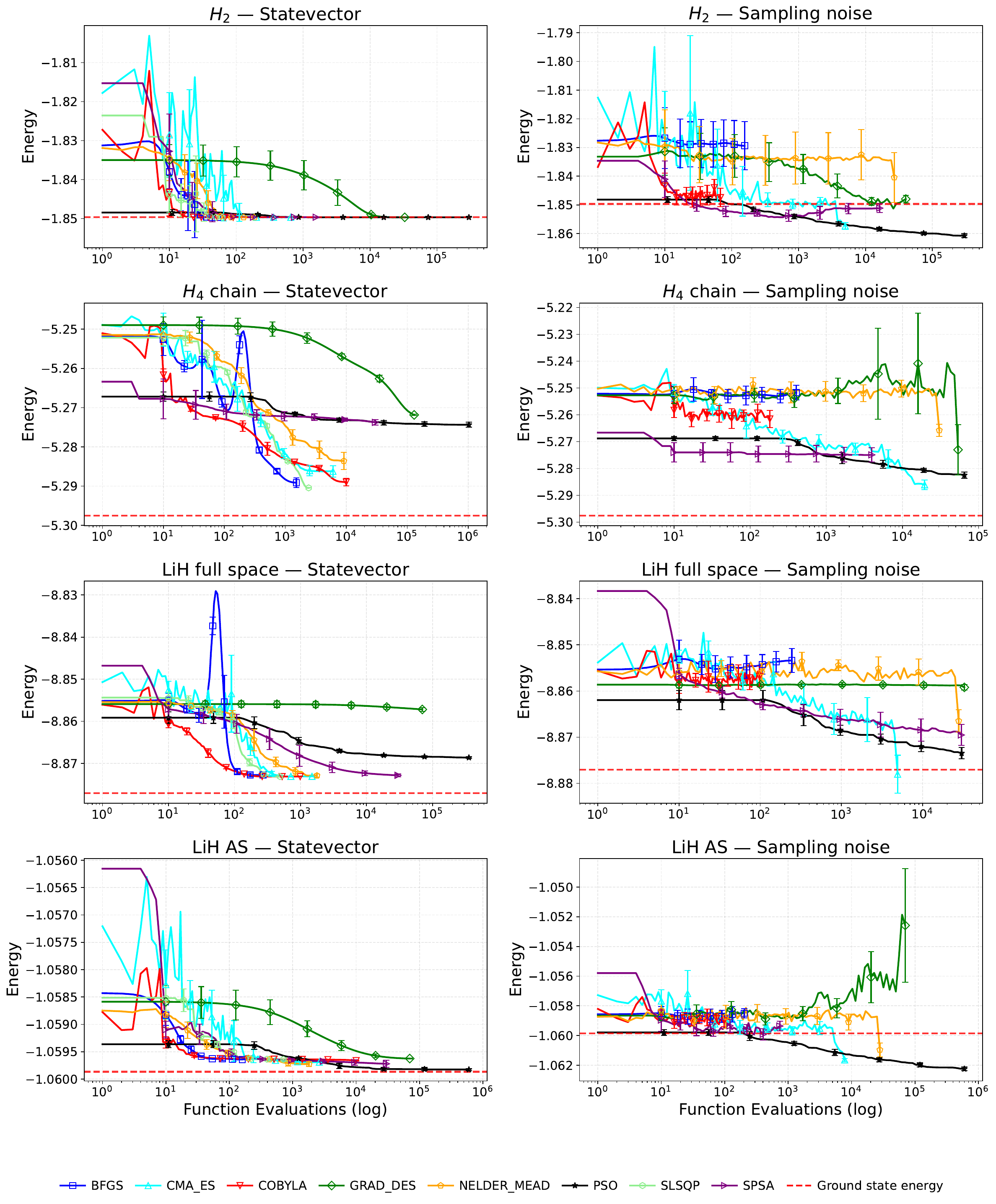}
    \caption{Convergence curves for \ce{H2}, \ce{H4}, and LiH in full and active space with random initialization starts using \ac{tvha}. Finite-shot sampling increases variance and erases the performance gap between deterministic and evolutionary optimizers. In the right panel we can see optimizers convergence to dip below ground state energies (red dashed line).}
    \label{fig:vha}
\end{figure}

Overall, the results across all four systems, summarized in Fig. \ref{fig:vha}, show a consistent evolution in optimizer behavior. In small, smooth landscapes (\ce{H2}), deterministic methods such as \ac{bfgs} excel, but as system size and noise increase (\ce{H4}, \ce{LiH}), stochastic and population-based optimizers \ac{cmaes}, PSO, and SPSA become dominant, achieving the lowest corrected errors and most stable convergence profiles. High-shot reevaluation proves indispensable, as apparent low-energy results from noisy runs often overestimate true accuracy. These findings emphasize a clear transition from curvature-driven optimization in noiseless low-dimensional systems to noise-resilient exploration strategies in larger, stochastic quantum simulations.

\section{Conclusion}

This work establishes a quantitative connection between measurement noise and optimizer instability in variational quantum algorithms. By benchmarking eight classical optimizers across molecular and condensed matter domains, we show that finite-shot sampling alone (without coherent or gate errors) is sufficient to destroy the deterministic structure of the variational landscape. The practical result is that gradient curvature becomes comparable to the noise amplitude, rendering gradient and Hessian information statistically meaningless.

Our main contribution is the systematic identification of where and why traditional optimizers fail. We demonstrate that deterministic gradient-based methods (\ac{bfgs}, \ac{slsqp}) lose reliability once curvature signals fall below the sampling variance; they transition from contracting toward a minimum to simple random diffusion. Conversely, adaptive population based algorithms such as \ac{cmaes} and \ac{ilshade} remain operational by implicitly averaging noise, which provides a general strategy for stochastic stability.

We also reveal a unique artifact to noisy quantum optimization: the winner’s curse, where statistical minima appear to fall below the true ground state energy. Our LiH study confirms that this bias can misleadingly suggest that noise is beneficial. This false advantage is removed by high-shot reevaluation, proving that the effect arises from estimator variance rather than physical noise assisted convergence. A statistically sound correction is to reevaluate elite individuals or track population means or trimmed means instead of relying on a single best sample.

From these analyzes, the reader can extract four practical takeaways for robust VQE optimization. First, structured ansatz as intrinsic noise mitigation. Compact, physically motivated designs such as \ac{tvha} maintain polynomially decaying gradients and stable curvature, reducing barren plateaus and noise sensitivity. Barren Plateaus are still a very active area of research, and further work will use a preferred, but more theoretically demanding, method for detecting BPs by analytically studying the scaling of the variance. For example, if $U(\boldsymbol{\theta})$ is a deep unitary circuit whose distribution of unitaries forms at least a 2-design over a Lie group, one can use representation theory to exactly evaluate the variances via Weingarten calculus. 

Second, evolutionary optimizers are designed to overcome multimodal, rotated, shifted, and compositionally complex benchmark functions, similar to the rugged and noisy landscapes characteristic of VQE cost surfaces. As shown in IEEE CEC benchmarks \cite{garcia2017since}, this robustness comes at the cost of increased function evaluations. However, this trade off can be advantageous in hybrid quantum–classical workflows. Methods such as \ac{ilshade} and \ac{cmaes} converged to nearly exact values even for the multimodal Hubbard model under heavy noise (64 shots). This behavior suggests a resource efficient optimization strategy where low shot evaluations are used throughout most of the process, with adaptive shot increases only near convergence to refine high-precision results.

Third, adaptive metaheuristics tend to have the best results in all our experiments. Algorithms that self-adapt their search parameters, such as \ac{cmaes} that updates its sampling distribution, and \ac{ilshade} that adaptively modifies its population size, crossover probability, and step size based on historical success, outperform static gradient-free, deterministic methods, and basic metaheuristics in the presence of noise. 

Fourth, dimensionality reduction as noise control. Simplifying the variational space through active space truncation or feature pruning directly lowers the effective noise floor, often improving optimizer reliability by an order of magnitude.

This study focused exclusively on finite-shot sampling noise and does not include any form of quantum hardware noise. In future work, we will study how different types of noise, including hardware decoherence, thermal relaxation, and readout errors, contribute to the problems analyzed in this study.

\section*{Acknowledgments}
Vojtěch Novák is supported by Grant of SGS No. SP2025/072, VSB-Technical University of Ostrava, Czech Republic. This work was supported by the Ministry of Education, Youth and Sports of the Czech Republic through the e-INFRA CZ (ID:90254 ). Martin Beseda is supported by Italian Government (Ministero dell'Università e della Ricerca, PRIN 2022 PNRR) -- cod.P2022SELA7: ''RECHARGE: monitoRing, tEsting, and CHaracterization of performAnce Regressions`` -- Decreto Direttoriale n. 1205 del 28/7/2023.

\section*{Data availability}
To support \ac{tvha} findings and ensure reproducibility, we have published the dataset containing both the computed results and the scripts used to generate them.
The dataset serves as a companion resource to this work and facilitates further research in hybrid quantum-classical architectures.
It is publicly available on Zenodo \cite{beseda2025results}.

Hubbard and Ising models simulations were conducted using a locally developed code, which is openly available for further examination and reproduction of results. The code, along with detailed Jupyter notebooks, can be accessed at \footnote{\url{https://github.com/VojtechNovak/VQAevolutionary}}. 

Further details are available through the corresponding author.

\appendix

\section{Experiment details}

We performed large-scale simulations of the \ac{tvha} on classical hardware to assess optimization performance in quantum chemistry settings. 
Four molecular systems \ce{H2}, \ce{LiH} (full space), \ce{LiH} (active space), and the \ce{H4} chain were optimized using eight classical algorithms: \ac{bfgs} \cite{liu1989}, \ac{cmaes} \cite{hansen2019pycma,Hansen2006CMAES}, COBYLA \cite{regis2011stochastic}, Gradient Descent, Nelder–Mead \cite{nelder1965simplex}, PSO \cite{eberhart1995particle}, \ac{slsqp} \cite{boggs1995sequential}, and SPSA \cite{Spall1998SPSA}. 
Each optimizer–molecule pair was evaluated in 20 independent runs (10 noiseless, 10 noisy) under random parameter initialization, yielding 640 total simulations with up to 10,000 iterations per run. 
Noisy configurations used 6144 measurement shots per function evaluation. 

Initial parameters were drawn independently from a uniform $[0,1)$ distribution using \texttt{numpy.random.rand}, and pseudorandom seeds were not fixed to preserve stochastic independence across runs. 
Each optimizer employed its standard termination criterion and default hyperparameters to ensure comparability. 
All implementations used a controlled Python environment combining \texttt{qiskit}~1.0.2 \cite{qiskit2024}, \texttt{qiskit\_nature}~0.7.2, \texttt{qiskit\_aer}~0.14.2, \texttt{pyscf}~2.6.0 \cite{pyscf2018}, \texttt{scipy}~1.15.2 \cite{2020SciPy-NMeth}, and \texttt{cma}~3.3.0 \cite{hansen2019pycma}, with full source and logs available on Zenodo~\cite{beseda2025results}.

For the Ising and Hubbard models, we have used optimizers: \ac{cmaes}, COBYLA, DEbest1bin and DEbest1exp both from \texttt{scipy} \cite{2020SciPy-NMeth}, SA Cauchy, iL-SHADE from \texttt{PyADE} \cite{pyade2025}, HS \cite{yang2009harmony} and SOS \cite{CHENG201498} from \texttt{mealpy} \cite{van2023groundwater,van2023mealpy}, iSOMA \cite{zelinka2023isoma}. For each setting, we performed five independent runs and report the mean results for both function evaluations (FEs) and convergence.

The \ac{cmaes} optimizer was further fine-tuned using the IRACE method. In contrast, iL-SHADE did not require additional tuning due to its self-adjusting nature, utilizing success-history-based parameter adaptation and a linear population size reduction scheme. For further details on hyperparameters, refer to \cite{Novak2025VQAmeta} or \footnote{\url{https://github.com/VojtechNovak/VQAevolutionary}}.

\bibliographystyle{splncs04}
\bibliography{apssamp}

\end{document}